\documentclass[10pt]{article}
\usepackage[a4paper,  margin=3cm]{geometry}
\usepackage{amssymb}
\usepackage{amsfonts}
\usepackage{amsmath}
\usepackage{graphicx}
\usepackage{amsfonts}
\usepackage{amssymb}
\usepackage{epstopdf}
\usepackage{color}
\usepackage[export]{adjustbox}
\usepackage{float}

\usepackage{tkz-graph}
\usepackage{tikz}
\usetikzlibrary{backgrounds}
\usetikzlibrary{trees,positioning,arrows}
\usetikzlibrary{calc}

\usepackage[section]{placeins}

\begin{document}
	\title{A stochastic epidemic model
		of COVID-19 disease}
	
	\author{Xavier Bardina {\footnote{Xavier Bardina and Carles Rovira are supported by the grant PGC2018-097848-B-I00.}
		}\\
		Departament de Matem\`atiques\\
		Universitat Auton\`oma de Barcelona\\
		08193-Bellaterra
		\and 
		Marco Ferrante \\
		Dipartimento di Matematica ``Tullio Levi-Civita''\\
		Universit\`a
		degli Studi
		di Padova \\
		Via Trieste 63, 35121-Padova, Italy
		\and Carles
		Rovira \\
		Departament de Matem\`atiques i Inform\`atica \\ 
		Universitat de Barcelona \\ 
		Gran
		Via 585, 08007-Barcelona}
	
	\date{}
	
	\maketitle
	
	\begin{abstract}
		To model the evolution of diseases with extended latency 
		periods and the presence of asymptomatic patients like COVID-19, we  
		define a simple discrete time stochastic 
		SIR-type epidemic model.
		We  include both latent periods as well as 
		the presence of quarantine areas,
		to capture the evolutionary dynamics of such diseases. 
	\end{abstract}
	

	
	

\section{Introduction}

There exists a wide class of mathematical models 
that analyse the spread of epidemic diseases, either
deterministic or stochastic, and 
may involve many factors such as infectious agents, mode of transmission,
incubation periods, infectious periods, quarantine periods, etcetera
(Allen 2003; Anderson and May 1991; Bailey 1975; 
Daley and Gani 1999; Diekmann et al. 2013).

A basic model of infectious disease population dynamics, consisting
of susceptible (S), infective (I) and recovered (R) individuals were first 
considered 
in a deterministic model by Kermack and McKendric (1927). 
Since then, various epidemic deterministic models have been developed,
with or without a time delay (see e.g McCluskey 2009 and 
Huang et al. 2010).
At the same time, many stochastic models 
have been considered: discrete time models
(see e.g Tuckwell and Williams 2007 ; Oli et al. 2006), 
continuous time Markov chain
models and diffusion models (see e.g. Mode and Sleeman 2000).
The models obtained in these three categories are of increasing mathematical complexity and allow to study several aspects of the epidemics. 

Even if the discrete-time models are the simplest ones, they may 
help to better define the basic principles of the contagion and to avoid 
the constraints due to the own definitions of the more sophisticated models.

In this paper we will adapt a simple SIR-type model proposed by 
Ferrante et al. (2016) and we will divide
the population into several classes to better describe the evolution
of the COVID epidemic.
Here we will need to include latency periods, the presence of asymptomatic 
patients and different level of isolation. 
To model the evolution of the epidemic, we will describe the evolution
of every single individual in the population, modelling the probability
on every day to be infected and, once infected, the exact evolution of its
disease until the possible recovery or the death. 
The construction of the theoretical model is carried out in Section 2,
where we are able to compute the probability of contagion  
and an estimate of the basic reproduction number $R_0$.
Then, in Section 3 we answer to five research questions by using
a simulation of the evolution of the disease. The use of the 
simulation is justified by the complexity of the model, that prevent to
carry out any further exact computation.
We are able to see that to stop the epidemic is fundamental to
start early with a severe quarantine and that a late starting date or a more
soft quarantine makes this procedure almost useless.
Moreover, to determine the quarantine it is very important to know 
the level of infectivity of the asymptomatic, since the more 
infectious they are, the more important is the quarantine.
Finally, as expected,
the group immunity plays a very important role to prevent the development 
of the disease.

\section{A SIR type discrete-time stochastic epidemic model}

To model the evolution of epidemics, Tuckwell and Williams (2007)
proposed a simple stochastic SIR-type model based on a discrete-time 
Markovian approach, later generalized by Ferrante et al. (2016) with a SEIHR model.
These models, despite their simplicity, are very
unrealistic to catch the characteristic of the COVID-19 disease and
for this reason
in this paper we introduce a more complex system, 
that we call SEIAHCRD, that better describes this new disease.

Assume that the population size is fixed and equal to $n$, and that the time
is discrete, with the unit for the duration of an epoch one day.
Every individual, marked by an integer between $1$ and $n$, 
belongs to one of the following 8 classes:
\begin{itemize}
	\item
	the class $S$ includes the individuals susceptible to the disease
	and never infected before; 
	\item 
	the class $E$ includes the individuals in a latency period, 
	i.e. individuals that have been infected but that are still 
	not infectious or sick;
	\item 
	the class $I$ includes the infectious individuals, that are not yet sick, but that will develop later the disease;
	\item 
	the class $A$ includes the infectious individuals, that are not yet sick, but that will NOT develop later the disease, usually referred as Asymptomatic;
	\item 
	the class $H$ includes the infectious individuals, that are sick, 
	but with light symptoms and therefore at home quarantine;
	\item
	the class $C$ includes the infectious individuals, that are sick
	and with severe symptoms, and most of the time are hospitalized;
	\item
	the class $D$ includes the deceased individuals;
	\item
	the class $R$ includes the recovered individuals.
\end{itemize}
Assuming that we start at time $0$,
we will define for any individual $i\in\{1,\ldots,n\}$ 
the family of stochastic processes 
$Y_\Xi^{i}=\{ Y_\Xi^{i}(t), t=0,1,2,\ldots\} $, such that 
$Y_\Xi^{i}(t)=1$ if the individual $i$ at time $t$
belongs to the class $\Xi$,  where $\Xi$ is equal to
$S,E,I,A,$ $H,C,D$ or $R$, and $0$ otherwise.
In this way, the total number of individuals in the
class $\Xi$ at time $t\ge 0$ will be denote by
$Y_\Xi(t)$ and will be equal to 
$\sum_{i=1}^{n} Y_\Xi^{i}(t)$.

Let us now fix the main assumption on the evolution of the epidemic.
At time $0$ all the individuals are in $S$, but one in class
$I$ or $A$ and that the evolution of the contagion follows these rules: 
\begin {enumerate}
\item {\em Daily encounters}: each individual $i$, over $(t,t+1]$, 
will encounter a number of other individuals equal to $N_i(t)$
which we will assume to be a deterministic value;
\item {\em Contagion probability}: if an individual 
who has never been diseased up to and including time $t$, encounters an 
individual in $(t,t+1]$ who belongs to the class $I$ or $A$,
then, independently of the results of other encounters, 
the encounter results in transmission of the disease 
with probability $q_I$ and $q_A$, respectively.
\item {\em Permanence in the classes}:
any individual, but one, starts from class S.
Once infected he/she moves to class $E$ and so on according
to the graph below.
The time spent in the classes $E$, $I$ and $A$
are of $r_E$, $r_I$ and $r_A$ consecutive days, respectively.
These values can be considered deterministic or stochastic. 
Any individual who enters the class $H$
remains in this class for 
$r_{HC}$ consecutive days with probability $\alpha$ 
or for $r_{HR}$ consecutive days with probability $1-\alpha$.  
Any individual who enters the class $C$ 
remains in this class for $r_{CD}$ consecutive days 
with probability $\lambda$ or for $r_{CR}$ consecutive days with 
probability $1-\lambda$. 
As before, the values of these four numbers of consecutive days can be 
considered deterministic or stochastic.
To conclude, we assume that the individuals once in class $R$ reamin
there forever, the same as for the class $D$.
\item {\em Transitions between the classes}:
we assume that any individual can moves between the classes according
to the following graph
\begin{center}
	\begin{tikzpicture}[->, >=stealth', auto, semithick, node distance=2cm]
	\tikzstyle{every
		state}=[fill=white,draw=black,thick,text=black,scale=1]
	
	\node (A)                     {$S$};
	
	\node (B)[right of=A] {$E$};
	
	\node (C)[above right of=B] {$I$};
	
	\node (D)[below right of=B] {$A$};
	
	\node (E)[right of=C] {$H$};
	
	\node (F)[right of=E] {$C$};
	
	\node (G)[right of=D] {$R$};
	
	\node (H)[right of=F] {$D$};
	
	\path
	
	(A) edge     node{$\beta$}     (B)
	
	(B) 
	edge     node{$\mu$}     (C)
	edge     node[anchor=east]{$1-\mu$}   (D)
	
	(C) 
	edge    node{$1$}     (E)
	
	(D) edge                node{$1$}     (G)

	(E) edge     node{$\alpha$}    (F)
	edge     node[anchor=east]{$1-\alpha$}     (G)

	(F)    
	edge   node[anchor=west] {$1-\lambda$}     (G)
	edge   node{$\lambda$}     (H)
	
	;
\end{tikzpicture}
\end{center}

Here $\beta, \mu, \alpha$ and $\lambda$ denotes the transition probabilities
and the transitions occur at the end of the permanence time spent by the
individual in the previous class.
Note that $\mu, \alpha, \lambda$ are parameters that depends only on the
specific nature of the disease, while $\beta$ depends on this and 
the number of individuals in the classes $I$ and $A$.

Any individual, once infected with probability $\beta$, follows one
of the four paths described here:
\begin{enumerate}
\item he/she transits through the states $E, I, H, C, D$, where 
he/she remains,
respectively, for $r_E$, $r_I$, $r_{HC}$ and $r_{CD}$ days, after 
which he/she dies and moves to class $D$.
\item he/she transits through the states $E, I, H, C, R$, 
where he/she remains, respectively, for $r_E$, $r_I$, $r_{HC}$ 
and $r_{CR}$ days, after which he/she becomes immune and
moves to class $R$.
\item he/she transits through the states $E, I, H,  R$, where 
he/she remains, respectively, for $r_E$, $r_I$ and  $r_{HR}$ days, after which he/she becomes immune and moves to class $R$.
\item he/she transits through the states $E, A,  R$, where 
he/she remains, respectively, for $r_E$ and  $r_{A}$ days, after which 
he/she becomes immune and moves to  class $R$.
\end{enumerate} 
It is immediate to see that the probability to follow any of these four paths
is equal to, respectively,
\[
\mu \alpha \lambda \quad , \quad \mu \alpha (1-\lambda)
\quad , \quad \mu (1-\alpha) \quad , \quad (1-\mu) \ .
\]

\end{enumerate} 

In order to evaluate the probability $\beta$ of contagion at time $t$
of an individual in class $S$, we will start
defining the probabilities of meeting an individual
in the classes $S, E, I, A$ and $R$ as equal. 
Note that the individuals in classes 
$H$ and $C$ are in total
quarantine and that it is not possible to meet them and 
that the individuals in 
class $D$ are removed from the system.
So, we will deal with three possibly different encounter probabilities:
\begin{itemize}
\item $p_I$ the probability of meeting 
an individual
belonging to the class $I$;
\item
$p_A$ the probability of meeting 
an individual belonging to class A;
\item $p_S$ the probability of meeting an individual 
that is not infectious, that is he/she 
belongs to the classes $ S, E$ or $R$. 
\end{itemize}
Assuming that the probability of meeting any individual 
is uniform and independent from the above defined classes, we
define these probabilities as
\begin{eqnarray}
\label{p123}
p_I&=&
\frac{y_I}{n-y_H-y_C-y_D-1},\\ \nonumber
p_A&=&\frac{ y_A}{n-y_H-y_C-y_D-1},\\
\nonumber 
p_S&=&1-p_I-p_A 
\end{eqnarray}
when $y_H+y_C+y_D<n-1$, while $p_I=p_A=0, p_S=1$ when $y_H+y_C+y_D=n-1$.
In the above formulas,
$y_I, y_A, y_H, y_C$ and $y_D$ denote the number of individuals in classes $I, A, H, C$ and $D$, respectively.

Denoting by 
$j_{iI}, j_{iA}$ 
the number of meetings of the $i$-th individual at time $t$ with individuals in the classes 
$I$ and $A$, respectively,
the probability to meet this proportion of individuals is
\begin{equation*}
\sum_{j_{iI},j_{iA}=1}^{N_i(t)} \frac{N_i(t)!}{j_{iI}!  j_{iA}! (N_i(t)-j_{iI}-j_{iA})!} p_{I}^{j_{iI}} p_{A}^{j_{iA}}p_S^{N_i(t)-j_{iI}-j_{iA}},
\end{equation*}
where  $N_i(t)$ denotes the daily encounters of the individual $i$.
We can easily derive the probability of contagion
\begin{equation*}
p_{j_{iI}+j_{iA}}=1-((1-q_I)^{j_{iI}} (1-q_A)^{j_{iA}})
\end{equation*}
where $q_I$  and $q_A$  denote the probability of transmission of the specific disease for individuals 
in classes $I$ and $A$, respectively, which are usually different.
Then the probability of contagion at time $t+1$ of a single individual is equal to
\begin{equation*}
\begin{split}
\beta_i &=\sum_{j_{iI},j_{iA}=1}^{N_i(t)}p_{j_{iI}+j_{iA}}\frac{N_i(t)!}{j_{iI}!  j_{iA}! (N_i(t)-j_{iI}-j_{iA})!} p_{I}^{j_{iI}} p_{A}^{j_{iA}}p_S^{N_i(t)-j_{iI}-j_{iA}}\\
&=1-\sum_{j_{iI},j_{iA}=1}^{N_i(t)}\frac{N_i(t)!}{j_{iI}!  j_{iA}! (N_i(t)-j_{iI}-j_{iA})!} ((1-q_I)p_{I})^{j_{iI}} 
((1-q_A)p_{A})^{j_{iA}}\\
& \ \ \ \ \times p_S^{N_i(t)-j_{iI}-j_{iA}}.
\end{split}
\end{equation*}
Substituting (\ref{p123}), we then get
$$
\beta_i=1-\Bigl(1- q_I p_I-q_A p_A\Bigr)^{N_i(t)} =
1-\Bigl(1- \frac{q_I y_I-q_A y_A}{n-y_H-y_C-y_D-1}\Bigr)^{N_i(t)}
$$
when $y_H+y_C+y_D<n-1$, while 
$\beta_i=0$
when $y_H+y_C+y_D=n-1.$
As done by Tuckwell and Williams (2007), we can use these formulas to simulate the spread of an 
epidemic under these general assumptions. Some results, similar to those presented in  Tuckwell and Williams (2007),
can be found 
in Ferrante et al. (2016).

To conclude, let us now consider the basic reproduction number $R_0$,
i.e. the expected number of secondary 
cases produced by an infectious individual 
during its period of infectiousness (see Diekmann et al. 1990).
In the present model, this refers to individuals that transits through
the classes $I$ or $A$.
Let us recall the threshold value of $R_0$, which establishes that 
an infection persists only if $R_0>1$.
As for the SIR-model proposed by Tuckwell and Williams 
we are not able to derive the exact explicit value of $R_0$, but
it is possible to extend their results, when $N_i(t) \equiv N$, for all $i$ and $t$, obtaining that
\[
R_0=\mu r_I q_I N + (1-\mu) r_A q_A N + O\biggl(\frac{1}{n-1}\biggr) \ .
\]
Note that the value of $R_0$ computed above is the 
basic reproduction number
at the beginning of the disease, when there is one infectious 
individual that has $N$ contacts in a population with $n-1$ 
susceptible inhabitants. 
When the disease is in an advanced development, keeping the total 
number of contacts $N$ some of them won't be with susceptible 
individuals and the number of cases produced by an infectious 
individual will be smaller.
At time $t$, let us call $S(t)$ the number of susceptible individuals 
(class $S$) and $X(t)$ the number of individuals removed from the
population (members of the classes $H$, $C$ and $D$). 
Then, given an individual in class $I$, set $Z_i(t)$ his number 
of contacts with susceptible individuals during the day $t$ and 
$Q_i(t)$ the number of 
individuals infected by $i$ at the end of day $t$. 
Clearly $Z_i(t)$ follows an hypergeomtric distribution with parameters  
$n-X(t)-1, S(t)$ and $N$ and, given $Z_i(t)=z, z \in \{0, \ldots,N\}$, 
it can be seen that $Q_i(t) \sim {\rm Bin}(z,q_I)$. 
Thus
\begin{eqnarray*}
{\rm E}(Q_i(t) | X(t),S(t))&= &{\rm E}\Big( {\rm E}(Q_i(t) | Z_i(t),X(t),S(t))| X(t),S(t)\Big) \\ &=&
{\rm E}( q_I Z_i(t)| X(t),S(t)) = q_I N \frac{S(t)}{n-1-X(t)}.
\end{eqnarray*}
Furthermore, using the same ideas in Ferrante et al. (2016), 
the number of secondary cases corresponding to this individual will be
\[
r_I q_I N \frac{S(t)}{n-1-X(t)}+ O\biggl(\frac{1}{n-1-X(t)}\biggr).
\]
Finally, the number of secondary cases produced by one 
arbitrary infectious individual (that can be in class $I$ or $A$) 
at time $t$ given $S(t)$ and $X(t)$ and that we will call $R_0(t)$, will be
\[
R_0(t)=
(\mu r_I q_I  + (1-\mu) r_A q_A) N \frac{S(t)}{n-1-X(t)}+ O\biggl(\frac{1}{n-1-X(t)}\biggr).
\]

\section{Application to a COVID19 type disease}

The COVID19 
is a highly contagious disease that has appeared at the end of 2019.
From all the information that is published every day, often contradictory, in the media, we can extract some properties of the disease. 
After the contagion, the virus remains 
in a latent state for 5-7 days before the individual 
became infectious. Then the individual can begin with symptoms in 3 days or he can continue asymptomatic, but probably infectious during two weeks. 
It is not known actually the number of asymptomatic people, 
but it will be probably bigger that the number of people with symptoms.  
About 75\% of persons with symptoms have just light symptoms that last for two 
weeks and after that they became recovered. The other 25\% get sever 
symptoms after a first period of 7-9 days with light symptoms. 
A 15\% of these patients with severe symptoms die in 4-6 days, 
while the other 85\% will recover after a period of 18-24 days. 

This disease can be described by the SEIAHRCD model defined before. 
According to the data above described, we can choose the deterministic 
values for the permanence in the classes  
$$
r_E=5, r_I=3, r_A=14, r_{HR}=14, r_{HC}=9, r_{CR}=20,  r_{CD}=4 
$$ 
and for the probabilities
$$
\mu=0.5, \alpha=0.25, \ \mbox{and} \ \lambda=0.15. 
$$ 
We also have to make an assumption on the ineffectiveness of 
the individuals when in the classes $I$ and $A$. We assume that
$$
q_I=0.2,  \ \mbox{and} \ q_A=0.05. 
$$ 
That is, we are assuming that asymptomatic individuals are less 
infectious than individuals who have symptoms.

$R_0$, the expected number of secondary 
cases produced by an infectious individual, is
equal with these parameters to
\[
R_0 = N \big(0.3 + 0.35\big) + O\biggl(\frac{1}{n-1}\biggr) \ .
\]
We see that the threshold value $R_0=1$
is obtained for $N=1.54$ and that
$R_0=3.5$ (one of the possible empirical values estimated on the
basis of real data) for $N=5.38$.

We study the spread of the contagion in
a closed population of 10000 inhabitants for a period 
of time of 180 days. 
We assume that an infectious individual (in class I) arrives to 
this healthy population and he/she remains there for 3 days. We 
also assume that $N_i(t)\equiv N$ for any $i$ and that therefore $\beta_i$ is
constant for any individual. 
This assumption is strong in the case of
a possible quarantine, but it is still reasonable.  

In this paper, since the analytical approach to this model  
is really complicated, 
we focus on the 
evolution of the disease by implementing a simulation using environment Maple. 
All the values presented here are the mean computed over 30 repetitions
of the simulation and we report also the confidence intervals.

We know very well that our model cannot explain exactly 
the COVID19 epidemic, since there are too many unknown 
aspects about this new disease, but
we believe that the study of the behaviour of our model 
can help to understand the COVID19 epidemic. 
More precisely, we will answer five Research Questions 
regarding the dynamics of the disease depending on 
some of the parameters involved. 
Particularly, we deal with 
(1) the importance of the number of contacts, 
(2)-(3) the effectiveness of a quarantine depending on
the moment it begins and on its duration, 
(4) the role of the asymptomatic depending of their level of infectiveness and 
(5) what happens with different levels of group immunization.

For each of these situations we study six quantities that 
we consider of major interest:
\begin{enumerate}
	
	\item Class D: The number of deceased, therefore individuals in class D, 
	after 180 days.
	
	\item Max class C: The maximum number of individuals in class C,
	i.e. with severe symptoms, in one day.
	
	\item Total class C: The sum of all the days spent by all the individuals 
	in the class C.
	
	\item Max new H: The maximum number of individuals that enter 
	in class H (people with symptoms) in one day.
	
	\item Day max new H: The day when it is reached the maximum number of individuals  
	that enter in class H  in one day.
	
	\item Prop. infected: The proportion of the population that has been infected after 180 days.

\end{enumerate}

For all these quantities we give a table with the mean and the 95\% confidence
interval for the mean in several situations. We also present a plot of 
one simulation of the number of deaths, the number of individuals in 
class C and the number of individuals that enter in class H each day 
for any of these situations along the 180 days.
Note that the number of deceased is considered assuming that the 
health service is able to give the same level of assistance 
whatever the number of patients is, but probably in the situations 
where the health service is more stressed the assistance will 
be worse and the number of deceased may increase.

\subsection{RQ1: How does the number of contacts N influence the spread of the disease?}

We consider the dependence on the number of contacts on the evolution of the disease when $N=10, 5, 4$ and $3$. 
\begin{table}[ht] 
	\caption{\small{Influence of number of contacts N}} 
	\centering 
	{\footnotesize
		\begin{tabular}{c c c c c } 
			\hline\hline 
			$N$ & 10 & 5 &  4 & 3  \\ [0.5ex] 
			\hline\hline 
			{\rm Class D}  & 189.16  & 176.03   & 168.56  & 159      \\
			&   (184.15,194.16)   & (162.97,189.08)   &  (151.69,185.45)   &      \\ \hline
			{\rm Max class C}  & 564,23    &  420,03   & 330,40    &    225  \\
			&  (557.04,571.49)   & (390.83,449.22)  &  (297.70,363.09)   &     \\ \hline
			{\rm Total class C}  &   12007,33  & 11524,13   & 10597,60     &  9571   \\
			&  (11847.58,12167.07)      & (11432.97,11595.28)   & (9553.21,11641.98)      &    \\ \hline
			{\rm Max new  H}  & 441.16     & 236.13  & 172,10    &   100  \\
			& (432.16,449.38)       & (219.84,252.41)    & (155.18,189.01)      &   
			\\ \hline
			{\rm Day max new H}   &  42.90       & 64.70    &79.53   &  98  \\
			&  (41.77,44.02)     & (63.11,66.28)   & (73.87,85.18)      &   \\ \hline
			{\rm Prop. infected}  & 0.9999      & 0.9489    &0.8852     & 0.8274    \\
			& (0.9998,0.9999)       & (0.8856,1)   & (0.7991,0.9712)      &   
			\\ \hline
			\hline 
		\end{tabular} 
	}
	\label{table:varN} 
\end{table} 

\noindent
Let us point out that when $N=5$ in one of the 30 simulations the epidemic did 
not go forward when $N=4$ it happened in two cases. On the other hand, 
when $N=3$ and with our initial conditions, only in 1 of the 30 simulations the 
disease went on. For this reason, in the table we only give the values of such case.

\begin{figure}[h!]
	\centering
	\includegraphics[width=0.30\textwidth]{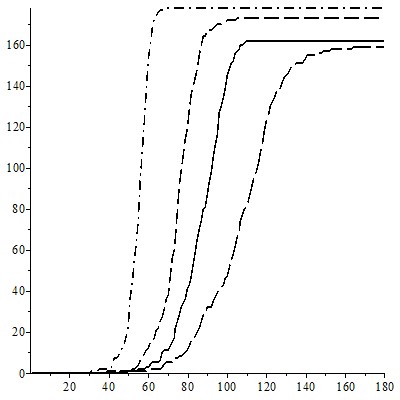}
	\includegraphics[width=0.30\textwidth]{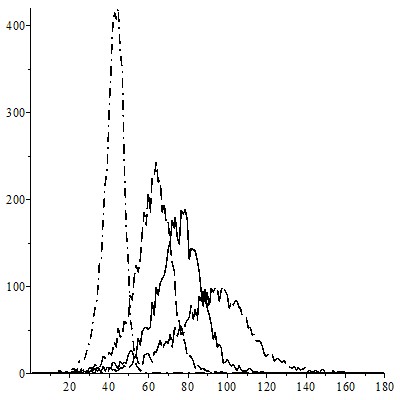}
	\includegraphics[width=0.30\textwidth]{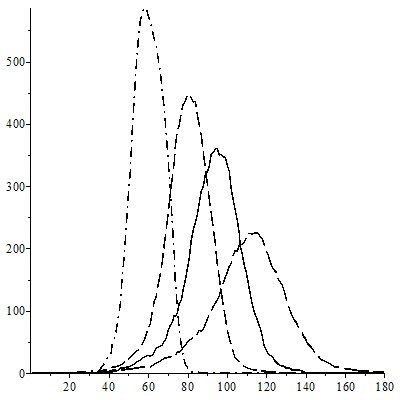}
	\caption{\small{Evolution of the number of deceased, 
			new individuals in H and the size of the class C when $N=10, 5, 4$ and $3$. }}
	\label{fig:mesvarN}
\end{figure}

\goodbreak
The number of deceased and the proportion of infected 
population are not very different between $N=4$ and $N=10$. The main 
difference consists in the velocity in the dissemination of the illness and so, 
the maximum of individuals in class C that can require hospitalization.

\subsection{RQ2: How does the beginning date of the 
	quarantine influence the spread of the disease?}

Let us check the effects of a quarantine on the evolution of the disease. 
We assume that $N=10$ at the beginning of the spread of the disease
and we consider three levels of quarantine, 
defined by the number of contacts $N=3$, $N=2$ or $N=0$ (the total quarantine).
We can also consider what happens depending on the moment that the quarantine starts:
\begin{itemize}
	\item when there are 10 deceased (first 10 individuals in class D),
	\item when there is the first deceased (first individual in class D),
	\item when there is the first individual with severe symptoms (first time the
	class C in not empty).
\end{itemize}

\subsubsection{Quarantine beginning after first 10 deceased.}
Let us recall that we assume that before the quarantine $N=10$.
We get
\begin{table}[ht] 
	\caption{\small{Influence of a quarantine of N at  10 deaths}} 
	\centering 
	{\footnotesize
		\begin{tabular}{c c c c c } 
			\hline\hline 
			$N$ & 10 & 3 &  2 & 0  \\ [0.5ex] 
			\hline\hline 
			{\rm Class D}  & 189.16  & 186.96   & 183.80  & 184.30      \\
			&   (184.15,194.16)   & (181.17,192.74)   &  (178.04,189.55)   &    (179.55,189.04)  \\ \hline
			{\rm Max class C}  & 564.23    &  564.60   & 557.46    &    560.66  \\
			&  (557.04,571.49)   & (557.04,571.41)  &  (551.60,563.31)   & (552.22,569.09)    \\ \hline
			{\rm Total class C}  &   12007.33  & 12013.86   & 11887.86   &  11960.53   \\
			&  (11847.58,12167.07)      & (11828.88,12198.83)   & (11775.82,11999.89)      & (11823.23,12097.82)   \\ \hline
			{\rm Max new  H}  & 441.16     & 439.73  & 440.66    &   441.90  \\
			& (432.16,449.38)       & (431.53,447.92)    & (434.07,447.24)      &   (435.30,448.49)
			\\ \hline
			{\rm Day max new H}   &  42.90       & 42.56    &43.93  &  43 \\
			&  (41.77,44.02)     & (41.70,43.41)   & (42.90,44.95)      & (42.08,43.91)  \\ \hline
			{\rm Prop. infected}  & 0.9999      & 0.9984    &0.9967     & 0.9934    \\
			& (0.9998,0.9999)       & (0.9983,9984)   & (0.9965,0.9968)      &   (0.9933,0.9934)
			\\ \hline
			\hline 
		\end{tabular} 
	}
	\label{table:varQ10} 
\end{table} 

\begin{figure}[ht]
	\centering
	\includegraphics[width=0.30\textwidth]{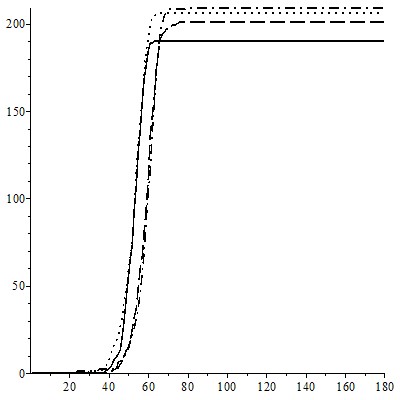}
	\includegraphics[width=0.30\textwidth]{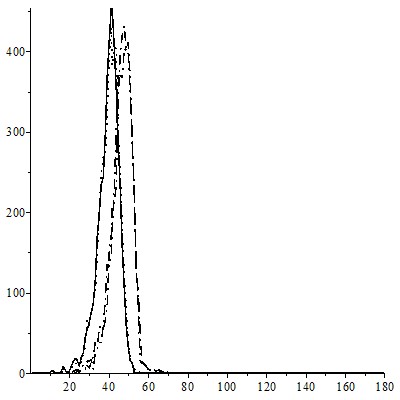}
	\includegraphics[width=0.30\textwidth]{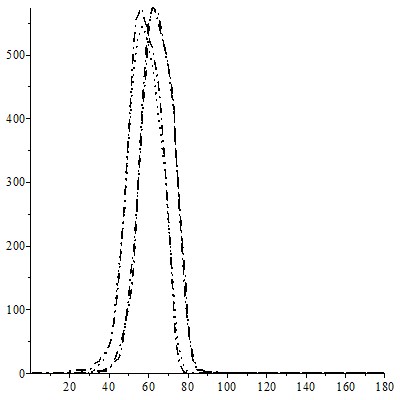}
	\caption{\small{Evolution of the number of  deceased, new individuals in H and the size of the class C  when the quarantine with $N=3, 2$ and $0$ begins with 10 deaths.}}
	\label{fig:mesvarQ10}
\end{figure}
It is clear that the efficacy of these quarantine, including the total 
quarantine, is almost null, since at this level of the disease 
the 99\% of the population has been infected (case $N=0$).
\newpage

\subsubsection{Quarantine beginning after the first deceased.}

\begin{table}[ht] 
	\caption{\small{Influence of a quarantine of N after the first death}} 
	\centering 
	{\footnotesize
		\begin{tabular}{c c c c c } 
			\hline\hline 
			$N$ & 10 & 3 &  2 & 0  \\ [0.5ex] 
			\hline\hline 
			{\rm Class D}  & 189.16  & 174.50   & 157.60  & 105.83      \\
			&   (184.15,194.16)   & (168.55,180.44)   &  (149.26,165.93)   &    (89.12,122,53)  \\ \hline
			{\rm Max class C}  & 564.23    &  433.23   & 386.80    &    365.40  \\
			&  (557.04,571.49)   & (401.39,465.06)  &  (339.66,433.93)   & (311.49,419.37)    \\ \hline
			{\rm Total class C}  &   12007.33  & 11176.66   & 10028.40  &  6860.66   \\
			&  (11847.58,12167.07)      & (10947.57,11405.74)   & (9453.42,10603.37)      & (5791.85,7929.46)   \\ \hline
			{\rm Max new  H}  & 441.16     & 347,50  & 343,06    &   367,73  \\
			& (432.16,449.38)       & (312.17,382.82)    & (299.08,387.03)      &   (324.30,407,15) 
			\\ \hline
			{\rm Day max new H}   &  42.90       & 42.70   &40.10  &  41.03 \\
			&  (41.77,44.02)     & (40.67,44.72)   & (38.41,41.78)      & (39.50,42.55)  \\ \hline
			{\rm Prop. infected}  & 0.9999      & 0.9401    &0.8412     & 0.5660    \\
			& (0.9998,0.9999)       & (0.9268,0.9533)   & (0.7975,0.8848)      &   (0.4769,0.6550)
			\\ \hline
			\hline 
		\end{tabular} 
	}
	\label{table:varQ1} 
\end{table}

\begin{figure}[ht]
	\centering
	\includegraphics[width=0.30\textwidth]{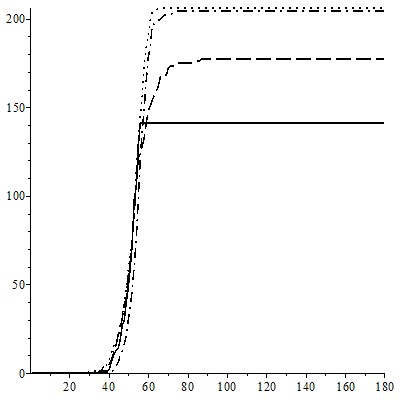}
	\includegraphics[width=0.30\textwidth]{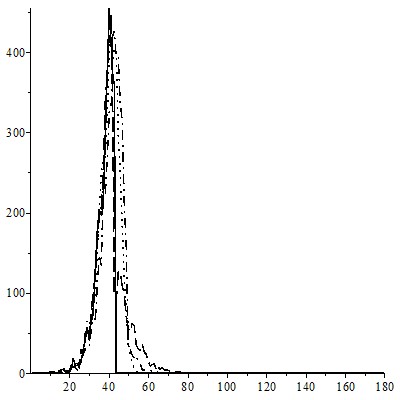}
	\includegraphics[width=0.30\textwidth]{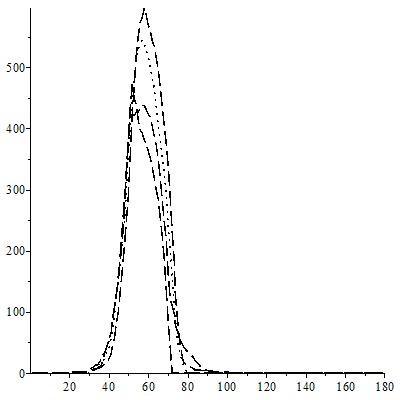}
	\caption{\small{Evolution of the number of deceased, new individuals in H and the size of the class C  when the quarantine with $N=3, 2$ and $0$ begins after the first death.}}
	\label{fig:mesvarQ1}
\end{figure}
In this case we can note the efficacy of a quarantine if it is rigorous, 
with a significant difference between $N=3$ and $N=2$. 
Note that in the total quarantine, there will be more that 100 
deceased since at this level of the disease more of the 50\% of 
the population has been infected (case $N=0$).

\subsubsection{Quarantine beginning after the first individual enters in class  C}

\begin{table}[ht] 
	\caption{\small{Influence of a quarantine of N after the first enter in C}} 
	\centering 
	{\footnotesize
		\begin{tabular}{c c c c c } 
			\hline\hline 
			$N$ & 10 & 3 &  2 & 0  \\ [0.5ex] 
			\hline\hline 
			{\rm Class D}  & 189.16  & 162.80   & 106.70  & 20.96      \\
			&   (184.15,194.16)   & (157.16,168.43)   &  (98.49,114,98)   &    (15.51,26.40)  \\ \hline
			{\rm Max class C}  & 564.23    &  272.93   & 130    &    75.50  \\
			&  (557.04,571.49)   & (259.34,286.51)  &  (104.41,155.58)   & (56.90,94,09)    \\ \hline
			{\rm Total class C}  &   12007.33  & 10382,53   & 6828,43  &  1337,87   \\
			&  (11847.58,12167.07)      & (10247.95,10517.10)   & (6430.56,7226.29)      & (1006.86,1668.87)   \\ \hline
			{\rm Max new  H}  & 441.16     & 149.36  & 96.96    &   117.13  \\
			& (432.16,449.38)       & (134.02,164.69)    & (68.38,125.53)      &   (92.21,143,04) 
			\\ \hline
			{\rm Day max new H}   &  42.90       & 50.33   &45.60  &  32.90 \\
			&  (41.77,44.02)     & (45.74,54.91)   & (35.97,55.22)      & (31.36,34.43)  \\ \hline
			{\rm Prop. infected}  & 0.9999      & 0.8662    &0.5709     & 0.1105    \\
			& (0.9998,0.9999)       & (0.8597,0.8726)   & (0.5388,0.6029)      &   (0.0839,0.1370)
			\\ \hline
			\hline 
		\end{tabular} 
	}
	\label{table:varQ1c} 
\end{table}

\begin{figure}[ht]
	\centering
	\includegraphics[width=0.30\textwidth]{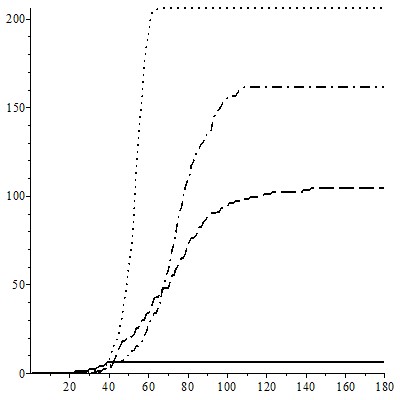}
	\includegraphics[width=0.30\textwidth]{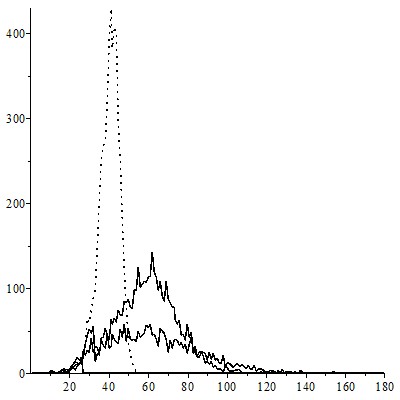}
	\includegraphics[width=0.30\textwidth]{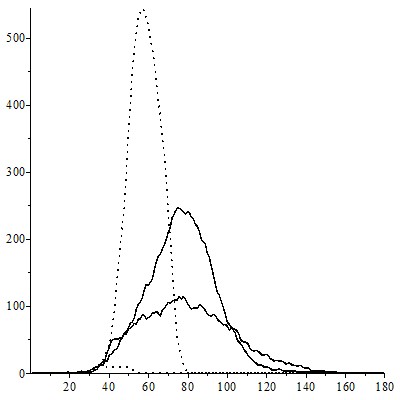}
	\caption{\small{Evolution of the number of  deceased, new individuals in H and the size of the class C  when the quarantine with $N=3, 2$ and $0$ begins after the first enter in C.}}
	\label{fig:mesvarQ1c}
\end{figure}

In this case the quarantine is effective. For $N=2$ we reduce the mortality 
of about 50\% and for $N=0$ the number of deceased is just 20. 
Note that when the first individual enters in C there 
are about 1000 individuals infected.

\newpage

\subsection{RQ3: How does the duration of the quarantine influence the spread of the disease?}

Let us consider now the duration of the quarantine. We consider 
quarantines, beginning at first individual in C, of duration 
20, 60 and 120 days. We also deal with two levels of quarantine, 
beginning from a number of contacts of $N=10$ we pass to a quarantine 
with $N=3$ or $N=1$.

\subsubsection{The case with $N=3$}
\begin{table}[H] 
	\caption{\small{Influence of a quarantine (from $N=10$ to $N=3$) of duration $k$ days after the first enter in C.}} 
	\centering 
	{\footnotesize
		\begin{tabular}{c c c c c } 
			\hline\hline 
			$k$ & 0 & 20 &  60 & 120  \\ [0.5ex] 
			\hline\hline 
			{\rm Class D}  & 189.16  & 186.50   & 172.26  & 160.43      \\
			&   (184.15,194.16)   & (181.19,191.80)   &  (165.68,178.83)   &  (155.63,165.22)    \\ \hline
			{\rm Max class C}  & 564,23    &  459.96   & 272.23   &    254.50 \\
			&  (557.04,571.49)   & (439.43,480.48)  &  (256.26,288.19)   &  (245.09,263.90)   \\ \hline
			{\rm Total class C}  &   12007,33  & 11869.33   & 11159.83    &  10319.40  \\
			&  (11847.58,12167.07)      & (11672.06,12066.59)   & (10971.99,11349.46)      &    (10249.01,10389.78)\\ \hline
			{\rm Max new  H}  & 441.16     & 370.5  & 146.73   &   131.03  \\
			& (432.16,449.38)       & (350.32,390.67)    & (133.95,159.50)      &  (124.94,137.11) 
			\\ \hline
			{\rm Day max new H}   &  42.90       & 55.03    &53.73   &  57.86 \\
			&  (41.77,44.02)     & (53.00,57.05)   & (48.46,58.99)      & (54.69,61.02)  \\ \hline
			{\rm Prop. infected}  & 0.9999      & 0.9982   &0.9274     & 0.8583    \\
			& (0.9998,0.9999)       & (0.9969,0.9999)   & (0.9156,0.9391)      &   (0.8527,0.8638)
			\\ \hline
			\hline 
		\end{tabular} 
	}
	\label{table:vark} 
\end{table} \begin{figure}[H]
	\centering
	\includegraphics[width=0.30\textwidth]{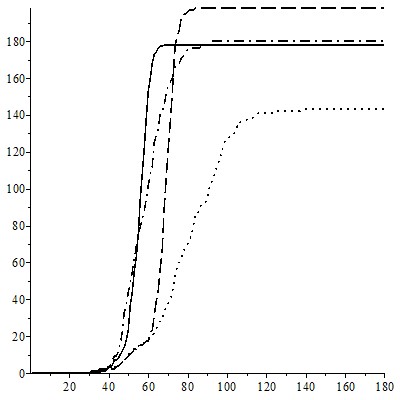}
	\includegraphics[width=0.30\textwidth]{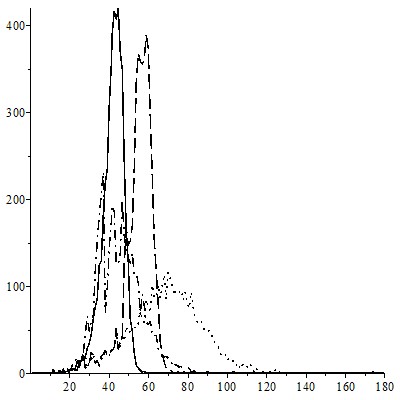}
	\includegraphics[width=0.30\textwidth]{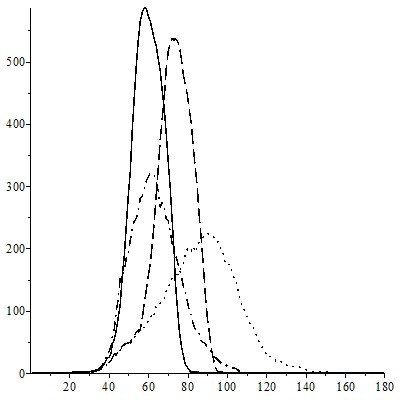}
	\caption{\small{Evolution of the number of  deaths, new individuals in H and the size of the class C when there is a quarantine of $k$ days.}}
	\label{fig:mesvark}
\end{figure}

When the quarantine is done with $N=3$, its effects in the number of deceased 
are not very important although we do a long quarantine of 120 days. 
Nevertheless, we are able to reduce the maximum in class C. 
To value the importance of the quarantine we need a more restricted quarantine.

\subsubsection{The case with $N=1$}

\begin{table}[h!] 
	\caption{\small{Influence of a quarantine (from $N=10$ to $N=1$) of duration $k$ days after the first enter in C.}} 
	\centering 
	{\footnotesize
		\begin{tabular}{c c c c c } 
			\hline\hline 
			$k$ & 0 & 20 &  60 & 120  \\ [0.5ex] 
			\hline\hline 
			{\rm Class D}  & 189.16  & 189.93   & 190.23  & 32.43      \\
			&   (184.15,194.16)   & (185.02,194.83)   &  (185.40,195.05)   &  (23.47,41.38)    \\ \hline
			{\rm Max class C}  & 564,23    &  487.40   & 429.60  &    72.00 \\
			&  (557.04,571.49)   & (465.44,509.35)  &  (401.45,457.74)   &  (45.05,98.94)   \\ \hline
			{\rm Total class C}  &   12007,33  & 12104.66  & 11951.36    &  2044.80  \\
			&  (11847.58,12167.07)      & (11954.15,12255.16)   & (11794.90,12107.81)      &    (1493.54,2596.05)\\ \hline
			{\rm Max new  H}  & 441.16     & 367,03  & 310,90   &   116,63  \\
			& (432.16,449.38)       & (347.37,386.68)    & (287.10,334.69)      &  (81.11,152.14) 
			\\ \hline
			{\rm Day max new H}   &  42.90       & 66.80    & 113.80   &  56.20 \\
			&  (41.77,44.02)     & (65.83,67.76)   & (106.08,121.51)      & (36.05,76.34)  \\ \hline
			{\rm Prop. infected}  & 0.9999      & 0.9993   &0.9982     & 0.2830   \\
			& (0.9998,0.9999)       & (0.9990,0.9995)   & (0.9973,0.9990)      &   (0.2047,0.3612)
			\\ \hline
			\hline 
		\end{tabular} 
	}
	\label{table:varkk2} 
\end{table} 

\begin{figure}[h!]
	\centering
	\includegraphics[width=0.30\textwidth]{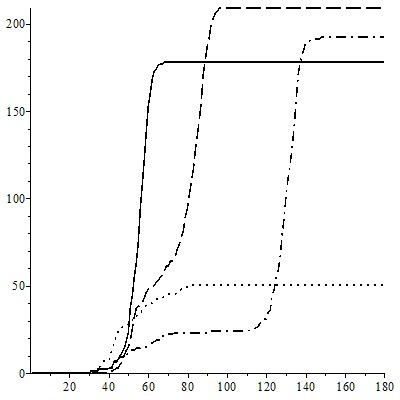}
	\includegraphics[width=0.30\textwidth]{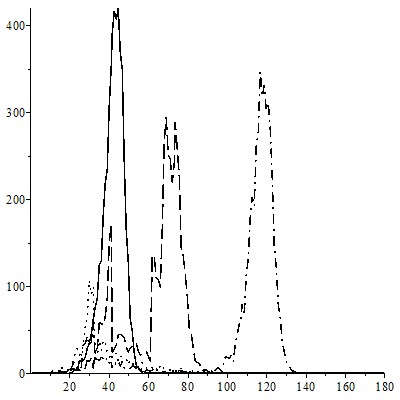}
	\includegraphics[width=0.30\textwidth]{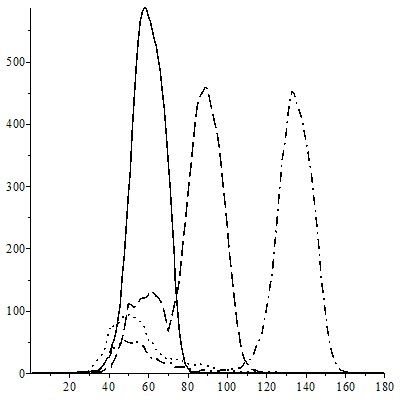}
	\caption{\small{Evolution of the number of  deaths, new individuals in H and the size of the class C when there is a quarantine of $k$ days.}}
	\label{fig:mesvark2}
\end{figure}

We can see here that with a strong quarantine ($N=1$) and with a long quarantine
($k=120$), the disease can be stopped. In the two other cases, $k=20$ or $k=60$ 
the final numbers of deceased are very similar, since after a short stop the 
disease returns to growth up arriving at the same levels of the initial 
outbreak and having two clear peaks.

\subsection{RQ4: How does the infectivity of asymptomatic individuals influence the spread of the disease?}

One of the main problems in the study of the COVID19 disease is the role of 
the asymptomatic. In our general framework we have supposed that 
the infectivity of an asymptomatic individual (class A) is one fourth 
of the infectivity of an infective individual that will have symptoms (class I),
that is, $q_A=0.05$ and $q_I=0.2$. Here we consider two other cases: 
in the first both probabilities are equal ($q_A=q_I=0.2$) and in the 
second case the asymptomatic individuals are not infectious ($q_A=0$). 
We also consider the case without quarantine and the case with 
quarantine (from $N=10$ to $N=3$) after the first individual in class C.

\subsubsection{The asymptomatics' role without quarantine}

\begin{table}[ht] 
	\caption{\small{Influence of $q_A$ without quarantine}} 
	\centering 
	{\footnotesize
		\begin{tabular}{c c c c  } 
			\hline\hline 
			$q_A$ & 0 & 0.05 &  0.2   \\ [0.5ex] 
			\hline\hline 
			{\rm Class D}  & 170.93  & 189.16  & 193.13      \\
			&  (153.62,188.16) &   (184.15,194.16)   & (187.07,199,18)        \\ \hline
			{\rm Max class C} & 458.66 & 564.23    &  607.76     \\
			&  (412.98,504.33) &  (557.04,571.49)   & (599.55,615.96)    \\ \hline
			{\rm Total class C}  & 11061.66 &   12007.33  & 11997.86     \\
			& (9967.41,12155.90)    &  (11847.58,12167.07)      & (11834.69,12161.02)      \\ \hline
			{\rm Max new  H}  & 302,73  & 441.16     & 642,76      \\
			& (272.29,333.16)  & (432.16,449.38)       & (631.81,653.70)         
			\\ \hline
			{\rm Day max new H}    &52.76  &  42.90       & 35.26 \\
			& (49.07,56.44)	&  (41.77,44.02)     & (34.63,35.88)         \\ \hline
			{\rm Prop. infected}  &0.9094 & 0.9999      & 1.0000          \\
			& (0.8209,0.9978) & (0.9998,0.9999)       & 1.0000   
			\\ \hline
			\hline 
		\end{tabular} 
	}
	\label{table:varA1} 
\end{table} 
\begin{figure}[ht]
	\centering
	\includegraphics[width=0.30\textwidth]{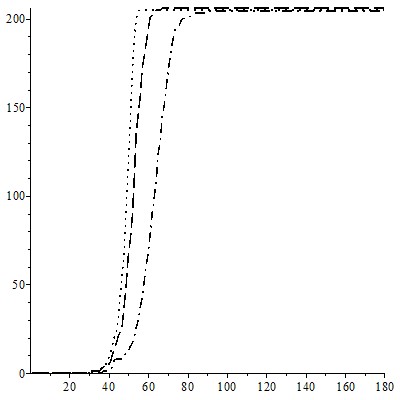}
	\includegraphics[width=0.30\textwidth]{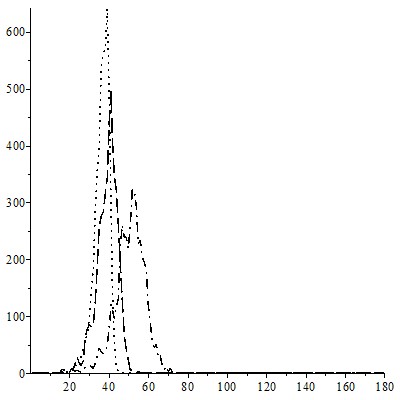}
	\includegraphics[width=0.30\textwidth]{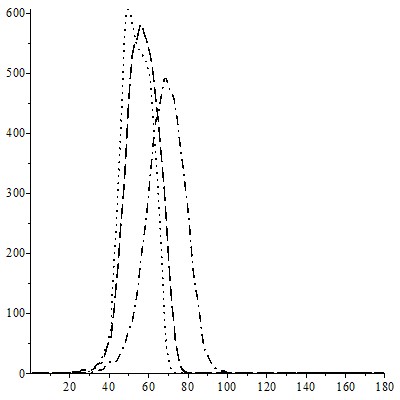}
	\caption{\small{Evolution of the number of  deaths, new individuals in H and the size of the class C  with different $q_A$ without quarantine.}}
	\label{fig:mesvarA1}
\end{figure}

We can see that the behaviour of the disease is very similar in 
the three cases and the main difference just consists in the velocity of its spread.

\subsubsection{The asymptomatics' role with quarantine}

\begin{table}[h!] 
	\caption{\small{Influence of $q_A$ with quarantine}} 
	\centering 
	{\footnotesize
		\begin{tabular}{c c c c  } 
			\hline\hline 
			$q_A$ & 0 & 0.05 &  0.2   \\ [0.5ex] 
			\hline\hline 
			{\rm Class D}  & 25.70  & 162.80  & 186.83    \\
			&  (19.43,31.96) &   (157.16,168.43)   & (181.99,191.66)        \\ \hline
			{\rm Max class C} & 39.00 & 272.93    &  537.93     \\
			&  (27.88,50.11) &  (259.34,286.51)   & (528.23,547.62)    \\ \hline
			{\rm Total class C}  & 1504.93 &   10382.53  & 11967.33     \\
			& (1102.08,1907.05)    &  (10247.95,10517.10)      & (11834.69,12161.02)      \\ \hline
			{\rm Max new  H}  & 132.43  & 149.36     & 452.33      \\
			& (116.75,148.10)  & (134.02,164.69)       & (409.95,494.70)         
			\\ \hline
			{\rm Day max new H}    &33.16  &  50.33      & 36.53\\
			& (30.62,35.69)	&  (45.74,54.91)     & (35.03,38.02)         \\ \hline
			{\rm Prop. infected}  &0.1353 & 0.8662      & 0.9990        \\
			& (0.1050,0.1655) & (0.8597,0.8726)       & (0.9979,1)  
			\\ \hline
			\hline 
		\end{tabular} 
	}
	\label{table:varA2} 
\end{table} 
\begin{figure}[h!]
	\centering
	\includegraphics[width=0.30\textwidth]{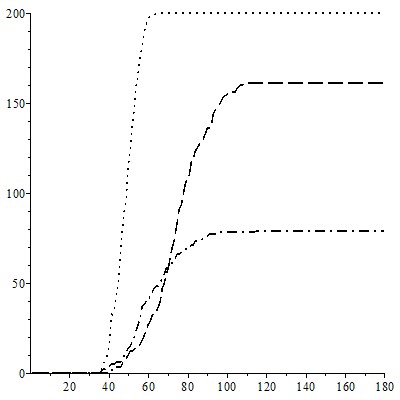}
	\includegraphics[width=0.30\textwidth]{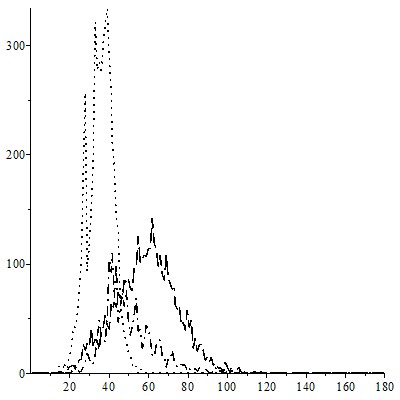}
	\includegraphics[width=0.30\textwidth]{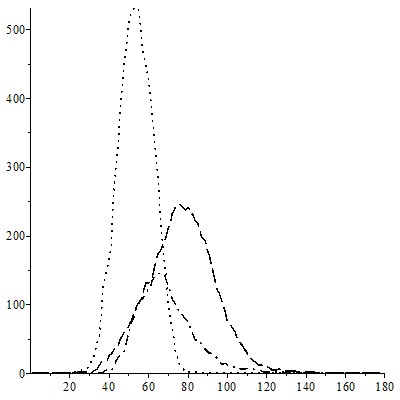}
	\caption{\small{Evolution of the number of  deaths, new individuals in H and the size of the class C  with different $q_A$ with quarantine.}}
	\label{fig:mesvarA2}
\end{figure}
We can see the importance of knowing the role of asymptomatics in implementing
quarantine. The more infectious the asymptomatic are, the more effective the
quarantine is.

\subsection{RQ5: How does the group immunity influence the spread of the disease?}

We consider finally the group immunity. We consider three cases, 
where the immunity group is of 50\%, 70\% and 90\%, respectively.
We can see that an immunity of the 50\% is not enough to control 
the disease since almost all the individuals without immunity 
become infected after the 180 days.

On the other hand, with an immunity of the 70\% the disease is 
well controlled. Moreover if the immunity regards the 70\% of the 
population, in 11 over 30 trials there aren't any deaths and in 
one of the trials the disease does not infect any person.

\begin{table}[H] 
	\caption{\small{Influence of group immunity}} 
	\centering 
	{\footnotesize
		\begin{tabular}{c c c c c } 
			\hline\hline 
			Level & 0\% & 50\% &  70\%  \\ [0.5ex] 
			\hline\hline 
			{\rm Class D}  & 189.16  & 90.53   & 18.00      \\
			&   (184.15,194.16)   & (87.07,93.98)   &  (12.78,23.21)      \\ \hline
			{\rm Max class C}  & 564,23    &  207.63   & 20.96  \\
			&  (557.04,571.49)   & (202.28,212)  &  (14.95,26.96)      \\ \hline
			{\rm Total class C}  & 12007.33  &5804.80  & 1124.06       \\
			&  (11847.58,12167.07)      & (5680.10,5929.49)   & (797.15,1450.96)      \\ \hline
			{\rm Max new  H}  & 441.16     & 120.76  & 12.40     \\
			& (432.16,449.38)       & (118.41,123.10)    & (9.06,15.73)      
			\\ \hline
			{\rm Day max new H}   &  42.90       & 61.30    & 94.26  \\
			&  (41.77,44.02)     & (58.94,63.65)   & (77.36,111.19)        \\ \hline
			{\rm Prop. infected}  & 0.9999      & 0.4848   &0.0990      \\
			& (0.9998,0.9999)       & (0.4842,0.4853)   & (0.0713,0.1266)     
			\\ \hline
			\hline 
		\end{tabular} 
	}
	\label{table:varG} 
\end{table} 

\begin{figure}[h!]
	\centering
	\includegraphics[width=0.30\textwidth]{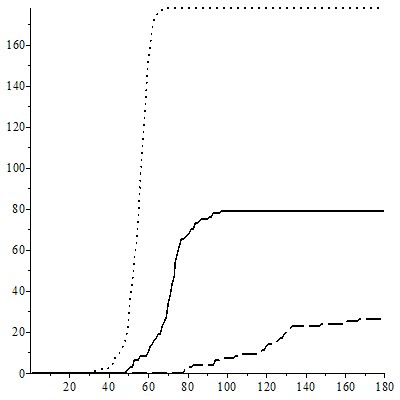}
	\includegraphics[width=0.30\textwidth]{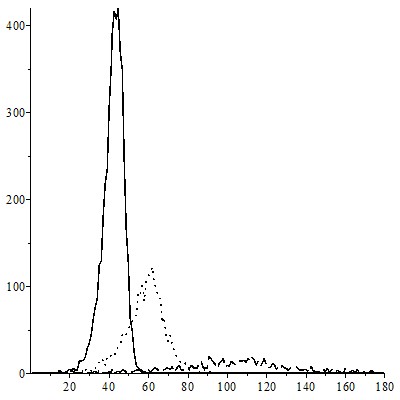}
	\includegraphics[width=0.30\textwidth]{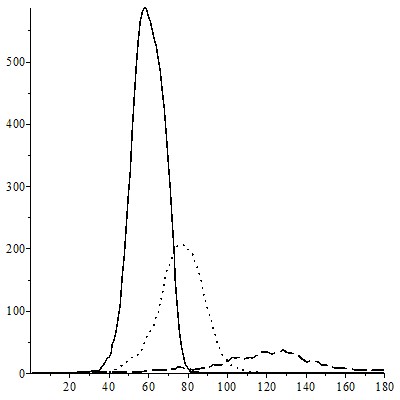}
	\caption{\small{Evolution of the number of  deaths, new individuals in H and the size of the class C when there is group immunity.}}
	\label{fig:mesvarG}
\end{figure}

If we consider an immunity of the 90\% only in 8 of the 30 trials there have been infected people with a maximum of 12 individuals infected and 
no one deceased.

\section{Conclusions}

Like any mathematical model, the model presented in this paper 
does not exactly describe COVID19 
disease, even if it shares with this most of its characteristic. 
Furthermore, for this new disease many aspects are still unknown, 
such as the level of infectivity of asymptomatic patients.

Our goal is to understand how a disease similar to the COVID19  
spreads over in a closed population and to answer to some specific
research questions.
What we have obtained can be summarized as follows:
\begin{enumerate}
	\item Reducing the number of contacts of each individual 
	the spread of the disease slows down,  
	the pressure on the health 
	system reduces, but we end up with a similar number of deceased.
	
	\item It is basic to start the quarantine at least when the first 
	severe patient is detected. 
	Waiting for the first deceased leads to many more additional deceased.
	
	\item To raise the quarantine it is very important to know the level of
	infectivity of the asymptomatic. The more infectious they are, 
	the more important is the quarantine.
	
	\item To stop the disease you must perform a strict and long quarantine 
	and you must start this as soon as possible.
	
	\item Group immunity is very important to prevent the development 
	of the disease.
\end{enumerate}

Even if most of the answers are the expected ones, we have obtained these
through a sound stochastic epidemic model, that despite of its simplicity
probably is able to catch most of the peculiarity of this disease.

We believe that a more sophisticated version of this model and more
elaborated simulations can allow us to answer to more 
complex questions, but probably it will be better to wait when a deeper
knowledge on this disease will be available.

\end{document}